\magnification=1200
\output={\plainoutput}
\input epsf
\baselineskip=\baselineskip

\newcount\pagenumber
\newcount\questionnumber
\newcount\sectionnumber
\newcount\appendixnumber
\newcount\equationnumber
\newcount\referencenumber
\newcount\subsecnumber

\global\subsecnumber=1

\def\ifundefined#1{\expandafter\ifx\csname#1\endcsname\relax}
\def\docref#1{\ifundefined{#1} {\bf ?.?}\message{#1 not yet defined,}
\else \csname#1\endcsname \fi}

\newread\bib
\newwrite\reffs

\newcount\linecount

\newcount\citecount
\newcount\localauthorcount 

\def\article{
\message{Defining article causes the references to be read in from  bib.tex}
\message{ and inserted directly into the article at the location of the use}
\message{of the command cite}
\message{For the obvious alternative style where the references are}
\message{written out to a separate file see the alternative command normalarticlestyle}
\def\eqlabel##1{\edef##1{\sectionlabel.\the\equationnumber}}
\def\seclabel##1{\edef##1{\sectionlabel}}                  
\def\feqlabel##1{\ifnum\passcount=1
\immediate\write\crossrefsout{\relax}  
\immediate\write\crossrefsout{\def\string##1{\sectionlabel.
\the\equationnumber}}\else \fi }
\def\fseclabel##1{\ifnum\passcount=1
\immediate\write\crossrefsout{\relax}   
\immediate\write\crossrefsout{\def\string##1{\sectionlabel}}\else\fi}
\def\cite##1{\immediate\openin\bib=bib.tex\global\citecount=##1
\global\linecount=0{\loop\ifnum\linecount<\citecount \read\bib 
to\temp \global\advance\linecount by 1\repeat\temp}\immediate\closein\bib}
\def\docite##1 auth ##2 title ##3 jour ##4 vol ##5 pages ##6 year ##7{
\par\noindent\item{\bf\the\referencenumber .}
 ##2, ##3, ##4, {\bf ##5}, ##6,     
(##7).\par\vskip-0.8\baselineskip\noindent{
\global\advance\referencenumber by1}}
\def\dobkcite##1 auth ##2 title ##3 publisher ##4 year ##5{
\par\noindent\item{\bf\the\referencenumber .}
 ##2, {\it ##3}, ##4, (##5).
\par\vskip-0.8\baselineskip\noindent{\global\advance\referencenumber by1}}
\def\doconfcite##1 auth ##2 title ##3 conftitle ##4 editor ##5 publisher ##6 
year ##7{
\par\noindent\item{\bf\the\referencenumber .}
##2, {\it ##3}, ##4,  {edited by: ##5}, ##6, (##7).
\par\vskip-0.8\baselineskip\noindent{\global\advance\referencenumber by1}}}

\def\normalarticlestyle{
\message{Defining normalarticlestyle causes references to be read in from bib.tex} 
\message{and written out to a file called reffs.tex.}
\message{}
\message{For the FINAL RUN simply paste in the contents of reffs.tex}
\message{where one wants the references to go}
\message{and comment out the usages of cite}
\message{If one also pastes in the file papermacro.tex}
\message{This then enables one to have a single tex file with everything in it.}
\message{This is desirable for giving somebody else a copy of an article}
\immediate\openout\reffs=reffs
\global\referencenumber=1
\def\eqlabel##1{\edef##1{\sectionlabel.\the\equationnumber}}
\def\seclabel##1{\edef##1{\sectionlabel}}                  
\def\feqlabel##1{\ifnum\passcount=1
\immediate\write\crossrefsout{\relax}  
\immediate\write\crossrefsout{\def\string##1{\sectionlabel.
\the\equationnumber}}\else \fi }
\def\fseclabel##1{\ifnum\passcount=1
\immediate\write\crossrefsout{\relax}   
\immediate\write\crossrefsout{\def\string##1{\sectionlabel}}\else\fi}
\def\z@{ 0pt}
\def\cite##1{\immediate\openin\bib=bib.tex\global\citecount=##1
\global\linecount=0{\loop\ifnum\linecount<\citecount \read\bib 
to\temp \global\advance\linecount by
1\repeat\immediate\write\reffs{\temp}\global\advance\referencenumber by1}
\immediate\closein\bib}
\def\docite##1 auth ##2 title ##3 jour ##4 vol ##5 pages ##6 year ##7{
\par\noindent\item{\bf\the\referencenumber .}
 ##2, ##3, ##4, {\bf ##5}, ##6,         
(##7).\par\vskip-0.8\baselineskip\noindent}
\def\dobkcite##1 auth ##2 title ##3 publisher ##4 year ##5{
\par\noindent\item{\bf\the\referencenumber .}
 ##2, {\it ##3}, ##4, (##5).
\par\vskip-0.8\baselineskip\noindent}
\def\doconfcite##1 auth ##2 title ##3 conftitle ##4 editor ##5 publisher ##6 
year ##7{
\par\noindent\item{\bf\the\referencenumber .}
##2, {\it ##3}, ##4,  {edited by: ##5}, ##6, (##7).
\par\vskip-0.8\baselineskip\noindent}}

\def\appendixlabel{\ifcase\appendixnumber\or A\or B\or C\or D\or E\or
F\or G\or H\or I\or J\or K\or L\or M\or N\or O\or P\or Q\or R\or S\or
T\or U\or V\or W\or X\or Y\or Z\fi}

\def\sectionlabel{\ifnum\appendixnumber>0 \appendixlabel
\else\the\sectionnumber\fi}

\def\beginsection #1
 {{\global\appendixnumber=0\global\advance\sectionnumber by1}\equationnumber=1
\par\vskip 0.8\baselineskip plus 0.8\baselineskip
 minus 0.8\baselineskip 
\noindent$\S$ {\bf \the\sectionnumber . #1}
\par\penalty 10000\vskip 0.6\baselineskip plus 0.8\baselineskip 
minus 0.6\baselineskip \noindent}

\def\subsec #1 {\bf\par\vskip8truept  minus 8truept
\noindent \ifnum\appendixnumber=0 $\S\S\;$\else\fi
$\bf\sectionlabel.\the\subsecnumber$ #1
\global\advance\subsecnumber by1
\rm\par\penalty 10000\vskip6truept  minus 6truept\noindent}

\def\beginappendix #1
{{\global\advance\appendixnumber by1}\equationnumber=1\par
\vskip 0.8\baselineskip plus 0.8\baselineskip
 minus 0.8\baselineskip 
\noindent
{\bf Appendix \appendixlabel . #1}
\par\vskip 0.8\baselineskip plus 0.8\baselineskip
 minus 0.8\baselineskip 
\noindent}

\def\no{\eqno({\rm\sectionlabel} .\the\equationnumber){\global\advance\equationnumber by1}}

\def\beginref #1 {\par\vskip 2.4 pt\noindent\item{\bf\the\referencenumber .}
\noindent #1\par\vskip 2.4 pt\noindent{\global\advance\referencenumber by1}}

\def\ref #1{{\bf [#1]}}


\normalarticlestyle
\advance\hsize by 2truept


\font\eightrm=cmr8
\font\sixrm=cmr6

\font\ninei=cmmi9
\font\eighti=cmmi8
\font\sixi=cmmi6
\skewchar\ninei='177 \skewchar\eighti='177 \skewchar\sixi='177

\font\ninesy=cmsy9
\font\eightsy=cmsy8
\font\sixsy=cmsy6
\skewchar\ninesy='60 \skewchar\eightsy='60 \skewchar\sixsy='60

\font\eightbf=cmbx8
\font\sixbf=cmbx6

\font\ninett=cmtt9
\font\eighttt=cmtt8

\hyphenchar\tentt=-1 
\hyphenchar\ninett=-1
\hyphenchar\eighttt=-1

\font\eightsl=cmsl8

\font\eightit=cmti8



\newskip\ttglue

\def\eightpoint{\def\rm{\fam0\eightrm}%
  \textfont0=\eightrm \scriptfont0=\sixrm \scriptscriptfont0=\fiverm
  \textfont1=\eighti \scriptfont1=\sixi \scriptscriptfont1=\fivei
  \textfont2=\eightsy \scriptfont2=\sixsy \scriptscriptfont2=\fivesy
  \textfont3=\tenex \scriptfont3=\tenex \scriptscriptfont3=\tenex
  \def\it{\fam\itfam\eightit}%
  \textfont\itfam=\eightit
  \def\sl{\fam\slfam\eightsl}%
  \textfont\slfam=\eightsl
  \def\bf{\fam\bffam\eightbf}%
  \textfont\bffam=\eightbf \scriptfont\bffam=\sixbf
   \scriptscriptfont\bffam=\fivebf
  \def\tt{\fam\ttfam\eighttt}%
  \textfont\ttfam=\eighttt
  \tt \ttglue=.5em plus.25em minus.15em
  \normalbaselineskip=9pt
  \let\sc=\sixrm
  \let\big=\eightbig
  \setbox\strutbox=\hbox{\vrule height7pt depth2pt width\z@}%
  \normalbaselines\rm}

\normalarticlestyle
\def\title{{\bf On non-$L^2$  solutions to the Seiberg--Witten equations}}

\headline={{\eightpoint \rm\ifnum\pageno=1\hfill\else\ifodd\pageno 
Adam, Muratori and Nash\hfill \title 
\else \title \hfill Adam, Muratori and Nash\fi\fi}}

\def\im{Im\,}

\def\longbar#1{\setbox1=\hbox{$#1$}
\setbox2=\vbox{\hrule width 0.8\wd1}
\raise0.5\ht1\hbox{${\lower\dp1\box2}\atop\box1$}}  
\def\mediumbar#1{\setbox1=\hbox{$#1$}
\setbox2=\vbox{\hrule width 0.6\wd1}
\raise0.5\ht1\hbox{${\lower\dp1\box2}\atop\box1$}}  

\def\dirac{\rlap/\partial}
\def\diraca{\rlap/\partial_A}

\par\vfill
\centerline{\title}
\vskip1.25\baselineskip
\centerline{by}
\vskip1.25\baselineskip
\centerline{Christoph Adam$^{*}$, Bruno Muratori$^{\dag}$ and 
Charles Nash$^{\dag}$}
\par\vskip\baselineskip
\noindent
Institut f\"ur Theoretische Physik$^*$\hfill Department of Mathematical Physics$^{\dag}$,
\par\noindent
Universit\"at Karlsruhe, \hfill National University of Ireland,   
\par\noindent
Germany.\hfill Maynooth,
\par\noindent
\null\hfill Ireland.
\par\vskip3\baselineskip
\noindent
{{\bf Abstract}: We show that a previous paper of Freund describing 
a solution to the Seiberg--Witten equations has a sign error rendering it a solution to a related but different set of equations. The non-$L^2$ 
nature of Freund's solution is discussed and clarified and we also 
construct a whole class of solutions to the Seiberg--Witten equations.}
\beginsection{Introduction}
With the introduction of the Seiberg--Witten equations \ref{1} 
there come  a wealth of results on four manifold theory and a new 
improved point  of view on Donaldson theory with an Abelian  gauge theory
supplanting a non-Abelian one---cf. \ref{2} for a review.   
\par
An important vanishing theorem of \ref{1}, reminiscent of the 
Lichernowicz--Weitzenb\"ock vanishing theorems, shows that there are no non-trivial 
solutions to the Seiberg--Witten equations on four manifolds 
with non-negative Riemannian scalar curvature. However one can have 
non-trivial solutions which are singular in some way---for example 
one could have a non-trivial solution which was not $L^2$: 
in \ref{3} Freund describes such a non-$L^2$ to the Seiberg--Witten 
equations on ${\bf R}^4$. Unfortunately a sign discrepancy  in \ref{3} means that the expressions  given there obey equations which differ from the Seiberg--Witten equations in a certain  sign. These other equations also admit $L^2$ solutions as well as non-$L^2$ ones and so Freund's equations 
are fundamentally different from the Seiberg--Witten equations.
\par
In $\S\,2$  we describe the Seiberg--Witten equations 
in a fairly explicit manner so as to expose notational   conventions
and matters of signs. In section $\S\,3$ we give the details concerning Freund's work and then in section $\S\,4$ we give an  $L^2$ solution 
of Freund's equations and a class of solutions to the Seiberg--Witten equations.
\beginsection{The Seiberg--Witten equations} 
If  $M$ is an oriented Riemannian four manifold with metric $g_{ij}$
then the data we need for the Seiberg--Witten equations are a 
$U(1)$ connection $A_i$ on  $M$ and a  local spinor field $M$. 
\par
If $F_{ij}$ is the curvature of $A_i$, so that its self-dual part $F^+_{ij}$ is given by $F^+_{ij}=1/2(F_{ij}+(\sqrt{g}/2)\epsilon_{ijkl} F^{kl})$, then the 
Seiberg--Witten equations are 
\eqlabel{\SWeqs}
$$\eqalign{F^+_{ij}&=-{i\over 2}\overline{M} \Gamma_{ij}M\cr
	\Gamma^i D_i M&=0\cr}\no		$$
where  $\Gamma_i$ are the gamma matrices\footnote*{\eightpoint Our conventions for the $\Gamma_i$ are: $\Gamma_0=\left(\matrix{0&I\cr I&0\cr}\right)$, 
$\Gamma_i=\left(\matrix{0&-i\sigma^i\cr i\sigma^i&0\cr}\right)$, 
and $\Gamma_5=-\Gamma_0 \Gamma_1 \Gamma_2 \Gamma_3$ where $\sigma^i$ are 
the Pauli matrices.}
satisfying $\{\Gamma_i,\Gamma_j\}=2g_{ij}I$, and $D_i$ and $\Gamma_{ij}$ 
are given  by
$$D_i=\partial_i+iA_i,\qquad 
\Gamma_{ij}={1\over2}\left[\Gamma_i,\Gamma_j\right]\no$$
In \ref{1} Witten also quotes the equations using the  
two component spinor formalism of L. Witten and Penrose, cf. \ref{4}, where 
the Gamma matrices make no explicit appearance. 
In this spinor form  the equations are
\eqlabel{\SWspinoreqs}
$$\eqalign{ {\cal F}_{A^\prime B^\prime}&={i\over 2}\left({\cal M}_{A^\prime}{\widetilde {\cal M}_{B^\prime}+
{\cal M}_{B^\prime}{\widetilde {\cal M}}_{A^\prime}}\right)\cr 
{\cal D}_{AA^\prime}{\cal M}^{A^\prime}&=0\cr}\no$$
\par
We now give a short summary of the relevant properties of the 
spinor formalism that we need here. 
\par
With a Riemannian metric of signature 
$(+,+,+,+)$ the 4 components of a $4$-vector $v_a\equiv(v_0,v_1,v_2,v_3)$ are represented by a $2\times2 $ matrix which is 
denoted by $v_{AA^\prime}$ and given by
$$v_{AA^\prime}={1\over\sqrt{2}}\left(\matrix{v_0+iv_3& iv_1+v_2\cr
                                              iv_1-v_2 &v_0-iv_3\cr }\right)\no$$

This expression for $v_{AA^\prime}$ can be written as a 
linear combination of what are known as the Infeld--van der Waerden matrices $g^a_{AA^\prime}$  defined by
$$g^0_{AA^\prime}={1\over\sqrt{2}}\left(\matrix{1 &0\cr
                                    0 & 1\cr}\right),\quad
g^i_{AA^\prime}={i\over\sqrt{2}}\sigma^i,\;i=1,2,3
\no$$
where $\sigma^i$ are the usual Pauli matrices so that
$$\sigma^1=\left(\matrix{0&1\cr 1&0\cr}\right),\quad
  \sigma^2=\left(\matrix{0&-i\cr i&0\cr}\right),\quad
  \sigma^3=\left(\matrix{1&0\cr 0&-1\cr}\right)\no$$
Using the $g^a_{AA^\prime}$'s the linear combination mentioned above is
given by
$$v_{AA^\prime}=v_a \, g^a_{AA^\prime}\no$$ 
and more generally if we have a tensor $T_{a_1 a_2\ldots a_n}$ it becomes
 ${\cal T}_{A_1A_1^\prime A_2A_2^\prime\ldots A_nA_n^\prime}$ where
$$ {\cal T}_{A_1A_1^\prime A_2A_2^\prime\ldots A_nA_n^\prime}=T_{a_1 a_2\ldots a_n} \,g^{a_1}_{A_1A_1^\prime} g^{a_2}_{A_2A_2^\prime}\cdots g^{a_n}_{A_nA_n^\prime}\no$$
In this formalism spinor indices are raised and lowered with the matrix 
$\epsilon_{AB}$ defined by 
$$\epsilon_{AB}=\left(\matrix{0&1\cr -1&0\cr}\right)=\epsilon^{AB}\no$$
For example, for one spinor index, one can write 
\eqlabel{\indexmoving}
$$v^A=\epsilon^{AB}v_B ,\quad v_B=v^A\epsilon_{AB}\no$$ 
An involution can also be defined on a spinor $v^A$  
taking it to a spinor $\widetilde v^A$
defined by
\eqlabel{\involution}
$$v^A=\left(\matrix{\alpha\cr \beta\cr}\right),\quad 
\widetilde v^A=\left(\matrix{-\bar\beta\cr  \bar\alpha\cr}\right) \no$$
where bar means complex conjugate.
\par
Simplification occurs  if the tensor is  antisymmetric  such as the 
curvature tensor $F_{ij}$: in that case one can verify that 
its spinor version ${\cal F}_{A A^\prime B B^\prime}$ is a linear combination of 
$\epsilon_{A^\prime B^\prime}$ and $\epsilon_{AB}$. More precisely one 
finds that
$$F_{ij}\, g^i_{A A^\prime} g^j_{B B^\prime}={\cal F}_{A A^\prime B B^\prime}
={\cal F}_{AB}\epsilon_{A^\prime B^\prime} +{\cal F}_{A^\prime B^\prime} \epsilon_{AB}
\no$$ 
Moreover it also turns out that ${\cal F}_{A^\prime B^\prime}$ and 
${\cal F}_{AB}$
are the spinor projections of the self-dual and anti-self-dual parts of 
$F_{ij}$ respectively; i.e. one can check that
$$F^+_{ij}\, g^i_{A A^\prime} g^j_{B B^\prime}={\cal F}_{A^\prime B^\prime}\epsilon_{AB},\quad F^-_{ij}\, g^i_{A A^\prime} g^j_{B B^\prime}=
{\cal F}_{AB}\epsilon_{A^\prime B^\prime}\no$$
\par
Now we return to the Seiberg--Witten equations and carry out the 
translation from the conventional to the spinor form. 
Starting with the Dirac equation we write
$$M=\left(\matrix{\alpha\cr \beta\cr0\cr0}\right)\equiv\left(\matrix{{\cal M}^{A^\prime}\cr0\cr0\cr}\right)\no$$
and then the Dirac equation 
$$\Gamma^i D_i M=0\no$$
becomes 
$$\sqrt{2}g^i_{AA^\prime}D_i{\cal M}^{A^\prime}=0\no$$
which we rewrite as
$${\cal D}_{AA^\prime}{\cal M}^{A^\prime}=0,\quad\hbox { with \quad }{\cal D}_{AA^\prime}=
{1\over\sqrt{2}}\left(\matrix{D_0+iD_3 & i D_1+D_2\cr
i D_1-D_2 & D_0-i D_3\cr
}\right)\no$$
which is the desired form. Moving on to the other equation 
$$F^+_{ij}=-{i\over 2}\overline{M} \Gamma_{ij}M \no$$
we first display this equation in full as
\eqlabel{\conventionalSW}
$$\eqalign{&{1\over2}\left(
\matrix{0& F_{01}+F_{23}& F_{02}-F_{13}& F_{03}+F_{12}\cr
        -F_{01}-F_{23} &0 & F_{03}+F_{12}& F_{13}-F_{02}\cr
         F_{13}-F_{02}& -F_{12}-F_{03} & 0 & F_{01}+F_{23}\cr
         -F_{12}-F_{03}& F_{02}-F_{13}& -F_{01}-F_{23}&0\cr}\right) 
=\cr
&\cr
&{1\over2}\left(\matrix{0 &\bar\beta\alpha+\bar\alpha\beta&i\bar\beta\alpha-i\bar\alpha\beta&\vert\alpha\vert^2-\vert\beta\vert^2\cr
                -\bar\beta\alpha-\bar\alpha\beta&0&\vert\alpha\vert^2-
\vert\beta\vert^2&-i\bar\beta\alpha+i\bar\alpha\beta\cr
                -i\bar\beta\alpha+i\bar\alpha\beta&-\vert\alpha\vert^2+
\vert\beta\vert^2&0&\bar\beta\alpha+\bar\alpha\beta\cr
                -\vert\alpha\vert^2+\vert\beta\vert^2&i\bar\beta\alpha-i\bar\alpha\beta &-\bar\beta\alpha-\bar\alpha\beta&0\cr}  \right)\cr}\no$$
and then find that 
$$F^+_{ij}\,g^i_{AA^\prime}g^j_{BB^\prime}=-{i\over 2}\overline{M} \Gamma_{ij}M\,g^i_{AA^\prime}g^j_{BB^\prime}\no$$
becomes 
$${\cal F}_{A^\prime B^\prime}\epsilon_{AB}={\cal T}_{A^\prime B^\prime}\epsilon_{AB}\no$$
i.e. 
\eqlabel{\spinorSW}
 $${\cal F}_{A^\prime B^\prime}={\cal T}_{A^\prime B^\prime}\no$$ 
where 
\eqlabel{\spinorextraone}
$${\cal F}_{A^\prime B^\prime}={1\over2}\left(\matrix{iF_{01}+iF_{23}+F_{13}-F_{02}&-iF_{12}-iF_{03}\cr
-iF_{12}-iF_{03}&-iF_{01}-iF_{23}+F_{13}-F_{02}\cr}\right)\no$$
and
\eqlabel{\spinorextratwo}
$${\cal T}_{A^\prime B^\prime}={1\over2}\left(\matrix{
2 i \bar\alpha\beta&-i\vert\alpha\vert^2+i\vert\beta\vert^2\cr
-i\vert\alpha\vert^2+i\vert\beta\vert^2&-2 i \alpha\bar\beta}\right)\no$$
On can now readily inspect equations \docref{conventionalSW}
and \docref{spinorSW}, \docref{spinorextraone}, \docref{spinorextratwo}
and confirm that the conventional and the spinor form of the equations 
agree.
\par
Also we can compute the matrix of components 
$(i/2)({\cal M}_{A^\prime}{\widetilde {\cal M}_{B^\prime}+
{\cal M}_{B^\prime}{\widetilde {\cal M}}_{A^\prime}})$ and verify that it is 
equal to ${\cal T}_{A^\prime B^\prime}$. Doing this we find that, if we start with 
${\cal M}^{A^\prime}=(\alpha, \beta)$ and use \docref{indexmoving} and 
\docref{involution}, we obtain

$$\eqalign{{i\over 2}(({\cal M}_{A^\prime}{\widetilde {\cal M}_{B^\prime}+
{\cal M}_{B^\prime}{\widetilde {\cal M}}_{A^\prime}}) 
&={1\over 2}\left(\matrix{2i \bar\alpha\beta& -i
\vert \alpha\vert^2 +i\vert\beta\vert^2\cr
-i\vert \alpha\vert^2 +i\vert\beta\vert^2 & -2i\alpha\bar\beta\cr}\right)\cr
&={\cal T}_{A^\prime B^\prime},  \cr}\no$$
as it should. We now turn to the explicit Fermion and gauge field considered by Freund.
\beginsection{Freund's equations}
In \ref{3} Freund chooses
$$\eqalign{A_i={1\over 2r(r-z)}\left(\matrix{0\cr -y\cr x\cr 0\cr}\right),\quad\hbox{ and }\quad {\cal M}^{A^\prime}&=
{1\over 2r\sqrt{r(r-z)}}\left(\matrix{x-iy\cr r-z\cr}\right)\cr
\widetilde {\cal M}^{A^\prime}&=
{1\over 2r\sqrt{r(r-z)}}\left(\matrix{-(r-z)\cr x+iy\cr}\right)\cr}
\no$$
for which one readily verifies that
$${\cal D}_{AA^\prime}{\cal M}^{A^\prime}=0\no$$
so that ${\cal M}^{A^\prime}$ is indeed a zero mode.
\par
To check the other equation we compute the curvature and find 
that, if $F_{ij}=\partial_i A_j-\partial_j A_i$, one has
$$\eqalign{F_{0i}&=0, \quad F_{12}=-{z\over 2 r^3},\quad 
F_{13}={y\over 2 r^3},\quad F_{23}=-{x\over 2 r^3}
\cr
\Rightarrow {\cal F}_{A{^\prime}B^{\prime}}&={1\over 4 r^3}
\left(\matrix{y-ix& iz\cr iz & y+ix\cr}\right)\cr}\no$$
On the other hand one also finds that
$$\eqalign{{i\over 2}({\cal M}_{A^\prime}{\widetilde {\cal M}_{B^\prime}+
{\cal M}_{B^\prime}{\widetilde {\cal M}}_{A^\prime}}) 
&={1\over 4 r^3}
\left(\matrix{-y+ix& -iz\cr -iz & -y-ix\cr}\right)\cr
&=-{\cal F}_{A{^\prime}B^{\prime}}\cr}$$
so that ${\cal F}_{A{^\prime}B^{\prime}}\not=(i /2)({\cal M}_{A^\prime}{\widetilde {\cal M}_{B^\prime}+
{\cal M}_{B^\prime}{\widetilde {\cal M}}_{A^\prime}}) $
and Freund's equations are 
\eqlabel{\freundseqs}
$$\eqalign{ {\cal F}_{A^\prime B^\prime}&=-{i\over 2}\left({\cal M}_{A^\prime}{\widetilde {\cal M}_{B^\prime}+
{\cal M}_{B^\prime}{\widetilde {\cal M}}_{A^\prime}}\right)\cr 
{\cal D}_{AA^\prime}{\cal M}^{A^\prime}&=0\cr}\no$$
\par
The Seiberg--Witten's equations are known to admit  no
non-trivial regular $L^2$ solutions in flat space 
(or spaces of positive scalar curvature) so  
Freund was concerned to point out that his fields provide an example 
of a non-trivial  solution which is not $L^2$. Unfortunately, as we have seen, Freund's fields, though not $L^2$, are not solutions to 
the Seiberg--Witten equations. 
\par
Since Freund's fields are static and have a connection with $A_0=0$
it is natural to consider them in ${\bf R}^3$. We now do this  
letting  ${\bf A}=(A_1,A_2,A_3)$ be the connection in ${\bf R}^3$ and 
denote its curvature components by $\widehat F_{ab}$, $a,b=1..3$. 
We obtain thereby the three 
dimensional  Freund equations
\eqlabel{\freundreducedeqs} 
$$\eqalign{\widehat F_{ab}&=-\epsilon_{abc}\overline{M} {\sigma^c}M,\quad a,b=1..3\cr
             \diraca M&=0,\quad\hbox{where }
\diraca=i\sigma^a(\partial_a+iA_a)\cr}\no$$    
In similar fashion we could also have obtained the three dimensional 
Seiberg--Witten equations---cf. \ref{2}---and, as in four 
dimensions, these differ from Freund's only in the 
sign of the quadratic Fermion term. They are   
\eqlabel{\reducedSW}
$$\eqalign{\widehat F_{ab}&=\epsilon_{abc}\overline{M} {\sigma^c}M,\quad a,b=1..3\cr
             \diraca M&=0,\quad\hbox{where }\diraca=i\sigma^a(\partial_a+iA_a)\cr}\no$$   
There is also a vanishing theorem which does not allow non trivial solutions in flat space so that there are no regular $L^2$ solutions to 
the equations  \docref{reducedSW} in ${\bf R}^3$.  However 
there is no such restriction on the Freund's equations 
\docref{freundreducedeqs}. In  the next section we show how to 
construct examples of {\it singular} non-$L^2$ solutions to the 
Seiberg--Witten equations and {\it regular} solutions to 
 of Freund's equations which are $L^2$ in ${\bf R}^3$.
\beginsection{The Freund and Seiberg--Witten  equations in three dimensions}
First of all we simply note that Freund's equations 
 \docref{freundreducedeqs} (or indeed \docref{freundseqs}) admit  the
following regular solution which is $L^2$ in ${\bf R}^3$. 
$$\eqalign{{\cal M}^{A^\prime} &= {\sqrt{12}({\bf 1}+i{\vec\sigma\cdot \vec r})\over (1+r^2)^{3/2}}
\left( \matrix {1  \cr 0  \cr} \right)\cr
A_i &=-{3\over (1+r^2)^2}
\left( \matrix {2 x z -2y  \cr 2 y z +2x\cr
 1-r^2 + 2 z^2  \cr} \right)\cr}\no$$
as may be checked easily. 
\par
Finally we would like to display some (necessarily singular) solutions 
to the three dimensional Seiberg--Witten equations 
\docref{reducedSW}; they will also of course be solutions of the full  
Seiberg--Witten equations \docref{SWeqs} or \docref{SWspinoreqs}. 
In fact we construct a whole class of such 
solutions parametrised by an arbitrary holomorphic function.
\par
First we need some facts about the Dirac equation in \docref{reducedSW}. A spinor
$M$ that obeys the Dirac equation $\diraca M=0$ of \docref{reducedSW} 
 must obey the condition
$${\partial_a \Sigma^a} =0,\quad\hbox{where }
\Sigma^a=\overline{M}\sigma^a M \no$$
 The connection $A_i$  in the Dirac equation  can be expressed  in 
terms of the zero mode $M$ by writing \ref{5}
\eqlabel{\connformula}
$$\eqalign{A_i &=-{1\over \sqrt{\Sigma^a \Sigma^a} }
\left({1\over2}\epsilon_{ijk}\partial_j
\Sigma_k +\im \overline{M} \partial_i M \right)\cr
&= -{1\over 2}\epsilon_{ijk}\left(\partial_j 
\ln \sqrt{\Sigma^a\Sigma^a}\right) N_k
-{1\over 2}\epsilon_{ijk}\partial_j  N_k -
\im \overline{\widehat M}
\partial_i \widehat M\cr
\hbox{where }\quad N^a&={ \Sigma^a\over \sqrt{\Sigma^b \Sigma^b}}, \quad \hbox{and }\widehat M ={M\over \sqrt{\overline{M}M}} \cr}\no$$ 
\par
Now, if $\chi$ is the complex variable
$$\chi={x+iy\over r^2}\no$$
and $G\equiv G(\chi,\bar\chi)$ is a function of $\chi$ and $\bar\chi$,
we obtain a new class of zero modes $M^G$ where 
$$\eqalign{M^G&=e^{G/2} M^0\cr
\hbox{and }\quad M^0&= {1\over r^3}\left(\matrix{z\cr x+iy\cr}\right) \cr}\no$$
The corresponding connection $A^G_i$ is found using formula 
\docref{connformula} and is given by
$$A^G_i=-{1\over 2}\epsilon_{ijk} \partial_j G N_k\no$$
We note that  the spinor $M^0$  is singular and non-$L^2$. For doing 
calculations it is also useful to note that $M^0$  
is a solution  of the free Dirac equation  
$$ \dirac M^0 =0\no$$
and that the  spin density for $M^G$ satisfies
$$\eqalign{\overline{M^G}\sigma^a M^G&=e^G \,\overline{M^0}\sigma^a M^0\cr 
\hbox{where }\quad \overline{M^0}\sigma^a M^0&= {N^a\over r^4} = 
{1\over 2i}{\bf (\partial\bar\chi)\times
(\partial\chi)}\cr}\no$$
The corresponding curvature $\widehat F^G_{ij}$ is 
 $$\eqalign{ \widehat F^G_{ij}&= -{\epsilon_{ijk}\over 2}\left[G_{,\chi} 
(\chi_{,kl}N_l +\chi_{,k}N_{l,l}
-\chi_{,ll}N_k - \chi_{,l}N_{k,l})+\right.\cr
& \quad G_{,\bar\chi} 
(\bar\chi_{,kl}N_l +\bar\chi_{,k}N_{l,l}
-\bar\chi_{,ll}N_k - \bar\chi_{,l}N_{k,l})-
\cr
&\left.\quad (G_{,\chi\chi}\chi_{,l}\chi_{,l} +G_{,\bar\chi \bar\chi}\bar\chi_{,l}
\bar\chi_{,l} +2G_{,\chi \bar\chi}\chi_{,l}\bar\chi_{,l}) N_k\right]
\cr}\no$$ 
After some tedious algebra we find that only the coefficient of
$G_{,\chi \bar\chi}$ is non-zero and that 
$$\chi_{,k}\bar\chi_{,k} = {2\over r^4}\no$$
and hence 
$$ \widehat F^G_{ij} ={2\epsilon_{ijk}\over r^4} G_{,\chi\bar\chi}N_k \no$$
But to have a solution of the Seiberg--Witten equations we must require that 
$$ \widehat F^G_{ij}=\epsilon_{ijk}\overline{M^G}\sigma^k M^G \no$$
and this means that
$$\eqalign{{2\over r^4} G_{,\chi\bar\chi}N_k&=
\overline{M^G}\sigma^k M^G\cr
\Rightarrow G_{,\chi\bar\chi}&={1\over 2} e^G\cr}
\no$$
as may be easily checked.
But this equation for $G$  is nothing other than the 
Liouville equation in the ``target space
coordinate'' $\chi$ with the ``wrong'' sign; that is to say that the sign  leads to the 
general singular solution
\eqlabel{\liouville}
$$G={\vert f^\prime(\chi)\vert^2\over (1 - f\bar f)^2}\no$$
where $f$ is an arbitrary holomorphic function of $\chi$. 
\par
Hence any pair 
$(M^G, A^G_i)$ with $G$ given by \docref{liouville} is a solution
the Seiberg--Witten equations \docref{reducedSW}. In fact, these 
solutions resemble the two-dimensional solutions of the Seiberg--Witten 
equations that were discussed in \ref{6}. 
Their solutions emerged as solutions to the same
Liouville equation \docref{liouville}, however the coordinate 
space variable $x_+ =x +iy$ appears rather than the target space variable 
$\chi$ used here.
\par
Finally we want to briefly describe the geometry of the spin density
term $\overline{M}\sigma^a M$. 
It is clearly rotationally symmetric around the $z$ axis and  
the integral curves of $\overline{M^0}\sigma^a M^0$  are circles that 
touch the $z$ axis at the point $z=0$. If we restrict to the $x-z$ plane, 
then these integral curves are the field lines of a dipole in two dimensions, 
and the vector field  $\overline{M^0}\sigma^a M^0$ restricted to that plane 
is  a scalar function  times the field of a dipole in two dimensions.
\par\vskip1.5\baselineskip
\noindent
{\bf Acknowledgment: } 
BM gratefully acknowledges financial support from the Training and 
Mobility of Researchers scheme (TMR no. ERBFMBICT983476).

\par\vskip2\baselineskip
\centerline{\bf References}
\par
\vskip0.5\baselineskip

{ \par \noindent \par \hangindent \parindent \indent \hbox to\z@ {\hss \fam \bffam \tenbf 1.\kern .5em }\ignorespaces  Witten E., Monopoles and four-manifolds, Math. Res. Lett., {\fam \bffam \tenbf 1}, 769--796, (1994).\par \vskip -0.8\baselineskip \noindent } 
{ \par \noindent \par \hangindent \parindent \indent \hbox to\z@ {\hss \fam \bffam \tenbf 2.\kern .5em }\ignorespaces  Donaldson S. K., The Seiberg--Witten equations and 4-manifold topology, Bull. Amer. Math. Soc., {\fam \bffam \tenbf 33}, 45--70, (1996).\par \vskip -0.8\baselineskip \noindent } 
{ \par \noindent \par \hangindent \parindent \indent \hbox to\z@ {\hss \fam \bffam \tenbf 3.\kern .5em }\ignorespaces  Freund P. G. O., Dirac monopoles and the Seiberg--Witten monopole equations, J. Math. Phys., {\fam \bffam \tenbf 36}, 2673--2674, (1995).\par \vskip -0.8\baselineskip \noindent } 
{ \par \noindent \par \hangindent \parindent \indent \hbox to\z@ {\hss \fam \bffam \tenbf 4.\kern .5em }\ignorespaces  Penrose R. and Rindler W., {\fam \itfam \tenit Spinors and Space-time vol. 1}, Cambridge University Press, (1984). \par \vskip -0.8\baselineskip \noindent } 
{ \par \noindent \par \hangindent \parindent \indent \hbox to\z@ {\hss \fam \bffam \tenbf 5.\kern .5em }\ignorespaces  Loss M. and Yau H., Stability of Coulomb systems with magnetic fields III. Zero energy bound states of the Pauli operator, Commun. Math. Phys., {\fam \bffam \tenbf 104}, 283--290, (1986).\par \vskip -0.8\baselineskip \noindent } 
{ \par \noindent \par \hangindent \parindent \indent \hbox to\z@ {\hss \fam \bffam \tenbf 6.\kern .5em }\ignorespaces  Nergiz S. and Sacioglu J., Liouville vortex and $\phi ^{4}$ kink solutions of the Seiberg--Witten equations, J. Math. Phys., {\fam \bffam \tenbf 37}, 3753--3759, (1996).\par \vskip -0.8\baselineskip \noindent } 
\bye